\documentclass[final,1p,times]{elsarticle} 
\usepackage{graphicx} 
\usepackage{amssymb} 
\usepackage{amsthm} 
\usepackage{lineno} 

\journal{Nuclear Physics A} 
\begin{document} 

\begin{frontmatter} 


\title{$\Upsilon$ production in d+Au collisions at STAR}

\author{Haidong Liu$^{a}$ for the STAR collaboration}

\address[a]{Department of Physics, University of California, 
 Davis, CA, 95616, USA}

\begin{abstract} 
We present the analysis of $\Upsilon\rightarrow e^{+}e^{-}$
production in d+Au collisions at $\sqrt{s_{NN}} = 200$ GeV from
the STAR experiment. Using higher level dedicated $\Upsilon$
triggers, STAR has sampled $32$ nb$^{-1}$ of integrated luminosity
in year 2008 d+Au run. The cross section is found as
$BR\times{({{d\sigma}\over{dy}})}_{y=0}^{\Upsilon(1S+2S+3S)}=35\pm4(stat.)\pm5(sys.)$
nb and it is consistent with NLO CEM prediction with
anti-shadowing effects. In addition we calculated the nuclear
modification factor $R_{dAu}=0.98\pm0.32(stat.)\pm0.28(sys.)$,
which suggests the $\Upsilon(1S+2S+3S)$ production follows
$N_{bin}$ scaling.
\end{abstract} 

\end{frontmatter} 



\section{Introduction}

Calculations based on lattice QCD predict that strongly
interacting systems at very high temperature lead to the
suppression of heavy quarkonium states due to Debye screening of
the color charges, providing a potential signature of Quark-Gluon
Plasma (QGP) formation in heavy ion collisions ~\cite{deba1}.
However, this simple picture is complicated by competing effects
that either reduce the yield, such as co-mover absorption
~\cite{peibero2,peibero3}, or enhance it, such as recombination
models ~\cite{peibero4,peibero5,peibero6}. With the low cross
section of the $\Upsilon$ family, the roles of absorption and
recombination are negligible. Lattice QCD studies of quarkonia
spectral functions suggest that while the $\Upsilon(3S)$ melts at
RHIC and the $\Upsilon(2S)$ is likely to melt, the $\Upsilon(1S)$
is expected to survive ~\cite{peibero7,peibero8}. Therefore, the
production of $\Upsilon$ family in pp, pA, and AA collisions is an
important tool to study the QGP properties ~\cite{deba2}. STAR has
successfully measured the $\Upsilon$ production in p+p
~\cite{qm06_proceeding} and Au+Au collisions
~\cite{qm08_proceeding}. It is important to study the $\Upsilon$
production in d+Au collisions to further understand the initial
and final state cold nuclear matter effects. In this article, we
report the preliminary results of the $\Upsilon$ cross section
measurement at midrapidity obtained with the STAR detector in d+Au
collision at $\sqrt{s_{NN}}=200$ GeV.
\section{Main Detectors}
The main detectors used in the STAR $\Upsilon$ measurements are
the TPC (Time Projection Chamber)~\cite{qm06_11} and the BEMC
(Barrel Electro-Magnetic Calorimeter)~\cite{qm06_10}. They cover a
pseudorapidity range of $\mid\eta\mid<1$ with full azimuthal
acceptance. The TPC is the tracking detector which measures the
particles momenta and energy loss ($dE/dx$) which provides
particle identification. The BEMC includes a total of 4800 towers. Each tower
covers approximately $0.05\times0.05$ in $\eta\times\phi$ space.
The BEMC can be used to trigger on high energy electron pairs thus
allowing us to maximize the sampled luminosity provided by RHIC.
Due to the removal of the Silicon Vertex Tracker (SVT) and Silicon
Strip Detector (SSD), a significant improvement of the signal to
background ratio for non-photonic electrons is achieved.

\section{$\Upsilon$ Trigger at STAR}
The STAR $\Upsilon$ trigger is based on a two-stage decision
comprising a level-0 (L0) hardware component and a level-2 (L2)
software component. The L0 trigger will be issued once a high
energy BEMC tower is measured. The L0 threshold is set to
$E_{T}\sim4.3$ GeV in Run8 d+Au collisions. The advantage of
triggering at such high energy is the added hadron rejection power
$e/h\sim100$ of the BEMC towers. The L2 trigger performs tower
clustering to reclaim energy leaked into neighboring towers. The
invariant mass is calculated as
$M_{ee}=\sqrt{2E_{1}E_{2}(1-\cos\theta)}$. Where $\theta$ is the
opening angle between clusters and $E_{1}$ and $E_{2}$ are the
energy of the clusters. Cuts are applied on $E_{1}>4.5$ GeV and
$E_{2}>3.0$ GeV, $\cos\theta<0$ and $6.5<M_{ee}<25$ GeV/$c^{2}$.

\section{Data Analysis and Results}
For d+Au collisions at $\sqrt{s_{NN}}=200$ GeV in 2008, the STAR
$\Upsilon$ trigger sampled $32$ nb$^{-1}$ of integrated
luminosity. The p+p equivalent integrated luminosity is $\sim12.5$
pb$^{-1}$, which is more than a factor of 2 larger than the sample
analyzed in the preliminary data reported in Ref
~\cite{qm06_proceeding} of the 2006 pp run. Electrons are
identified ~\cite{deba13} by selecting charged particle tracks,
where more than 22 out of 45 points are fitted and whose specific
ionization energy loss in the TPC is $-2<n\sigma_{e}<3$
($n\sigma_{e}$ is the normalized dE/dx and it is defined by
$n\sigma_{e}=\log(dE/dx\mid_{measure}/B_{e})/\sigma_{e}$, where
$B_{e}$ is the expected mean dE/dx of electrons and $\sigma_{e}$
is the dE/dx resolution of the TPC). The selected particle tracks
have to match to BEMC clusters that contain energy consistent with
the L0 and L2 trigger conditions. A cut on $0.7<E/p<1.3$ is
applied to further reject hadrons from the electron candidates,
where $E$ and $p$ are the particle's energy deposited in BEMC and
the momentum measured by the TPC, respectively. The $e^{+}e^{-}$
pairs are then combined to produce the invariant mass spectrum.
The combinatorial background is estimated using like-sign pairs.
The like-sign electron pairs are combined to form the invariant
mass spectrum and the geometric mean $2\sqrt{N^{++}N^{--}}$ is
used to calculate the combinatorial background, where $N^{++}$
($N^{--}$) is the number of the positive (negative) like-sign
electron pairs. The left panel of Fig. 1 shows the unlike-sign and
the like-sign invariant mass spectra. The right panel of Fig. 1
shows the transverse momentum reach for the unlike-sign and
like-sign electron pairs indicating that STAR has the capability
to measure the transverse momentum spectra of the $\Upsilon$
family up to $p_{T}\sim7$ GeV/$c$.
\begin{figure}
\begin{center}
\includegraphics[scale=0.7]{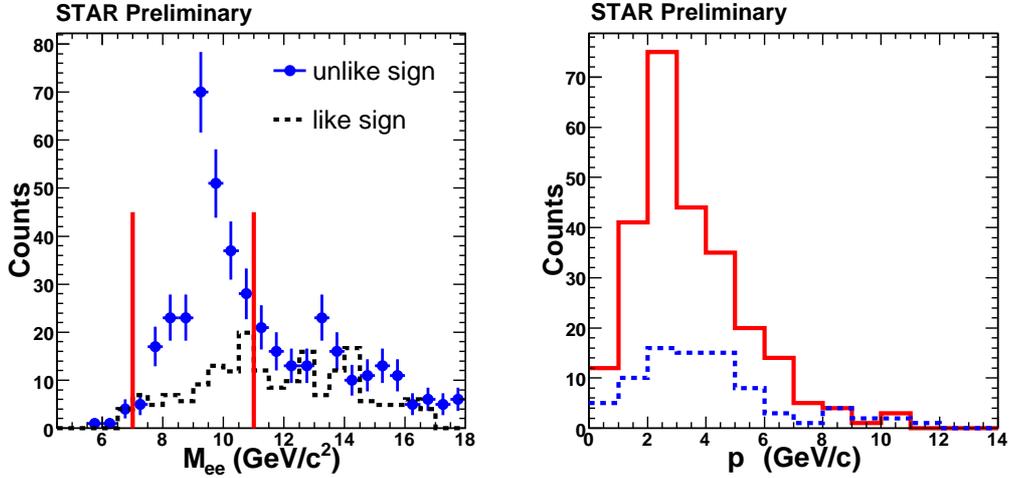}
\caption{Left panel: STAR $\Upsilon\rightarrow e^{+}e^{-}$ signal
and background in $\sqrt{s_{NN}}=200$ GeV d+Au collisions. The
solid symbols with statistical error bars are obtained by
combining the unlike-sign ($e^{+}e^{-}$) pairs. The dashed
histogram shows the like-sign background. Right panel: The
transverse momentum distributions for unlike-sign (solid line) and
like-sign (dashed line) $ee$ pairs.} \label{Tech_figure}
\end{center}
\end{figure}
The total yield is extracted by integrating the invariant mass
spectrum from 7 to 11 GeV/$c^{2}$. The invariant mass range used
for integration is the same as that for the STAR measurement of
p+p ~\cite{qm06_proceeding}. The integrated yield is
$172\pm20(stat.)$ and the signal significance is $\sim8\sigma$.
This is to-date, the strongest signal of $\Upsilon(1S+2S+3S)$
states at RHIC energies. The cross section is then calculated with
the formula:
\begin{equation}\label{xsection}
\sum_{nS=1}^3BR(nS)\times{({{d\sigma}\over{dy}})}_{y=0}^{\Upsilon(nS)}={{N}\over{dy\times\epsilon\times\int{Ldt}}}
\end{equation}
where $BR(nS)$ is the branching ratio fraction for
$\Upsilon(nS)\rightarrow e^{+}e^{-}$, $N=172\pm20(stat.)$ is the
raw yield, $dy=1.0$ is the rapidity interval, $\int{Ldt}=32$
nb$^{-1}$ is the integrated luminosity and
$\epsilon=\epsilon_{acc}\times\epsilon_{L0+L2}\times\epsilon^{2}_{TPC
reco}\times\epsilon^{2}_{eID cuts}\times\epsilon_{mass}$ is the
efficiency for reconstructing $\Upsilon$ family. $\epsilon_{acc}$
is the geometrical acceptance, $\epsilon_{L0+L2}$ is the L0 and L2
trigger efficiency, $\epsilon_{mass}$ is the efficiency of the
invariant mass cut, $\epsilon_{TPC reco}$ and $\epsilon_{eID
cuts}$ is the efficiency for single electron reconstruction and
the electron identification cuts, respectively. Each term and its
uncertainty is estimated through simulations and the total
efficiency is found to be $\epsilon=0.15\pm0.02$. The
cross-section at midrapidity in $\sqrt{s_{NN}}=200$ GeV d+Au
collisions is found to be
$BR\times{({{d\sigma}\over{dy}})}_{y=0}^{\Upsilon(1S+2S+3S)}=35\pm4(stat.)\pm5(sys.)$
nb. The contribution to the di-electron yield in the $\Upsilon$
mass region coming from Drell-Yan and $b\overline{b}$ is estimated
to be $\sim$10$\%$ based on ~\cite{vogt_dAu} and PYTHIA. The
detailed systematic uncertainty is under study.

We compare our midrapidity measurement with NLO
(next-to-leading-order) calculations the Color Evaporation Model
(CEM) ~\cite{vogt_dAu} in Fig. 2. The calculation includes the
anti-shadowing effect which is obtained from the EKS'98
parameterization ~\cite{EKS98}. It doesn't include absorption
effect. We also calculate the nuclear modification factor
$R_{dAu}=0.98\pm0.32(stat.)\pm0.28(sys.)$. The cross section in
p+p collisions is taken from STAR measurement
~\cite{qm06_proceeding}.
\begin{figure}
\begin{center}
\includegraphics[scale=0.55]{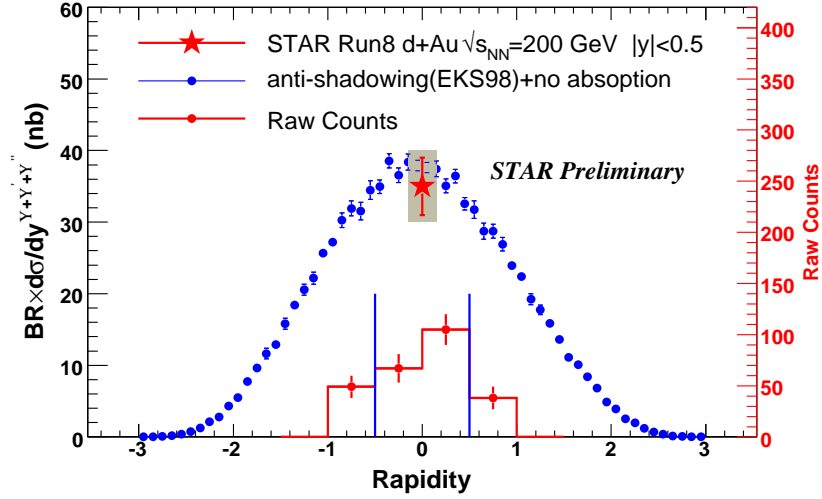}
\caption{The red star shows the measured
$BR\times{({{d\sigma}\over{dy}})}_{y=0}^{\Upsilon(1S+2S+3S)}$ at
midrapidity. The bar indicates the statistical error and the band
shows the systematic uncertainty. The cross section is compared
with the NLO CEM model prediction (blue solid circles), see text
for details. The raw yields vs. rapidity is shown by the red histogram at
the bottom with the statistical errors.} \label{Tech_figure}
\end{center}
\end{figure}
\section{Conclusion}
The STAR experiment made the first measurement of the
$\Upsilon(1S+2S+3S)\rightarrow e^{+}e^{-}$ cross section at
midrapidity in d+Au collisions at $\sqrt{s_{NN}}=200$ GeV. An
$8\sigma$ significant signal is observed and
$BR\times{({{d\sigma}\over{dy}})}_{y=0}^{\Upsilon(1S+2S+3S)}=35\pm4(stat.)\pm5(sys.)$
nb. The large acceptance of the detectors and the L0+L2 trigger
are essential for the success of the measurement. The cross
section is consistent with CEM predictions including
anti-shadowing. It is found that
$R_{dAu}=0.98\pm0.32(stat.)\pm0.28(sys.)$, which suggests the
$\Upsilon(1S+2S+3S)$ production follows the number of binary
collisions scaling in d+Au collision. With the current uncertainty
of $R_{dAu}$ we can not quantify the cold nuclear matter effect.
The uncertainties are dominated by the statistical error of the
$\Upsilon$ cross section measurement in p+p collision. In the RHIC
2009 p+p run, STAR has sampled 21 pb$^{-1}$ of integrated
luminosity and that will provide us a precise baseline for
$\Upsilon$ measurements.

\end{document}